\documentclass[twocolumn,10pt,eqno]{article}
\usepackage{amsmath,amssymb}

\newcommand{\gm}{\gamma}

\newcommand{\dlt}{\delta}

\newcommand{\tht}{\theta}

\newcommand{\vtht}{\vartheta}
\newcommand{\kp}{\kappa}

\newcommand{\sgm}{\sigma}
\newcommand{\Sgm}{\Sigma}
\newcommand{\vph}{\varphi}

\newcommand{\be}{\begin{equation}}
\newcommand{\ee}{\end{equation}}
\newcommand{\bea}{\begin{eqnarray}}
\newcommand{\eea}{\end{eqnarray}}
\newcommand{\eql}{\!\!\!&=\!\!\!&}

\newcommand{\defa}{\!\!\!&\equiv\!\!\!&}

\newcommand{\tl}[1]{\tilde{#1}}

\newcommand{\der}{\partial}
\newcommand{\dr}{\!\!d}
\newcommand{\hc}{{\rm h.c.}}
\newcommand{\ie}{{\it i.e.}}

\newcommand{\vev}[1]{\langle #1 \rangle}

\newcommand{\brkt}[1]{\left( #1 \right)}
\newcommand{\brc}[1]{\left\{ #1 \right\}}

\newcommand{\abs}[1]{\left| #1 \right|}

\renewcommand{\Re}{{\rm Re}}
\renewcommand{\Im}{{\rm Im}}

\newcommand{\cI}{{\cal I}}

\newcommand{\cL}{{\cal L}}

\newcommand{\cN}{{\cal N}}
\newcommand{\cO}{{\cal O}}

\newcommand{\cW}{{\cal W}}

\renewcommand{\ge}[2]{e_{#1}^{\;\;#2}}
\newcommand{\udl}[1]{\underline{#1}}

\newcommand{\lrder}{\stackrel{\leftrightarrow}{\partial}}

\newcommand{\NP}[1]{{\it Nucl.~Phys.}~{\bf #1}}
\newcommand{\PL}[1]{{\it Phys.~Lett.}~{\bf #1}}

\newcommand{\PR}[1]{{\it Phys.~Rev.}~{\bf #1}}
\newcommand{\PRL}[1]{{\it Phys.~Rev.~Lett.}~{\bf #1}}
\newcommand{\PTP}[1]{{\it Prog.~Theor.~Phys.}~{\bf #1}}

\newcommand{\JH}[1]{{\it JHEP}~{\bf #1}}

\begin{document}

\title{\bf Dynamical Radion Superfield in Five-dimensional Action\footnote{Talks 
 given by Y.Sakamura at PASCOS'05 (Geongju, Korea, May 30-June 4, 2005) 
 and at SUSY'05 (Durham, U.K., July 18-23, 2005).}}


\author{$\mbox{Hiroyuki Abe}^\dagger$ and $\mbox{Yutaka Sakamura}^\ddagger$ \\
  $\mbox{}^\dagger${\small\it Department of Physics, Kyoto University, 
  Kyoto, 606-8502, Japan} \\
  $\mbox{}^\ddagger${\small\it Department of Physics, Osaka University, 
  Osaka, 560-0043, Japan}}

\date{\begin{flushleft}\small {\bf Abstract.}
We clarify the radion superfield dependence of 
5D $\cN=1$ superspace action. 
The radion is treated as a dynamical field and appears in the action 
with the correct mode function. 
Our derivation is systematic and based on the superconformal 
formulation of 5D supergravity. 
We can read off the couplings of the dynamical radion superfield 
to the matter superfields from our result. 
The correct radion mass can be obtained by calculating the radion potential 
from our superspace action. \end{flushleft}}

\maketitle


\section{Introduction}
Five dimensional supergravity (5D SUGRA) compactified 
on an orbifold~$S^1/Z_2$ has been thoroughly investigated. 
Especially, the Randall-Sundrum model~\cite{RS} is attractive 
as an alternative solution to the hierarchy problem, 
and a large amount of research on this model has been done.  
In this model, the background geometry is a slice of the anti-de Sitter 
(AdS) spacetime and the metric has the form of 
\bea
 ds^2 \eql g_{\mu\nu}dx^\mu dx^\nu 
 = e^{-2ky}\eta_{mn}dx^m dx^n-dy^2 \nonumber\\ 
 \eql e^{-2kR\vtht}\eta_{mn}dx^m dx^n-R^2 d\vtht^2, 
\eea
where $\mu,\nu,\cdots=0,1,2,3,4$ and $m,n,\cdots=0,1,2,3$ are 
the 5D and 4D indices, and 
the coordinate of the fifth dimension is denoted as $y\equiv x^4$. 
The constant~$k$ is the AdS curvature and $R$ is the radius of the orbifold. 
The physical range of the extra space is $0\leq y \leq \pi R$. 
In the second line, we have changed the coordinate~$y$ 
to the dimensionless coordinate $\vtht\equiv y/R$. 

In such a brane-world model, the radius of the compactified extra dimension 
is generically a dynamical degree of freedom, {\it the radion}. 
In the original Randall-Sundrum model~\cite{RS}, the radius of the orbifold 
is undetermined by the dynamics and thus the radion is a massless field. 
Hence, it remains to be a dynamical degree of freedom 
in low energies and should be taken into account in the 4D effective theory. 
A naive way of introducing the radion mode into the theory is 
to promote the radius~$R$ to a 4D field~$r(x)$. 
Namely, the radion~$r(x)$ appears in the metric as~\cite{GW2,CGRT} 
\be
 ds^2 = e^{-2kr(x)\vtht}g^{(4)}_{mn}(x)dx^m dx^n-r^2(x)d\vtht^2, 
 \label{naive_radion}
\ee
where $g^{(4)}_{mn}$ is the 4D graviton. 
However, this is not an appropriate introduction of the radion. 
In fact, $r(x)$ in Eq.(\ref{naive_radion}) is not a mass-eigenstate. 
In the usual dimensional reduction procedure, we will expand the bulk fields into 
the infinite Kaluza-Klein (K.K.) modes. 
For example, a 5D field $B$ is mode-expanded as 
\be
 B(x,y)=\sum_n f_{(n)}(y)b_{(n)}(x). 
\ee
In order for $b_{(n)}(x)$ to be mass-eigenstates, 
we have to choose the mode-functions~$f_{(n)}(y)$ as solutions of 
the mode-equations, which are obtained from the linearized equation of motion 
for $B$. 
Otherwise, $b_{(n)}(x)$ are no longer mass-eigenstates 
and we cannot simply drop the higher K.K. modes after the heavy modes 
are integrated out. 
Note that the correct mode-function of the radion, which is proportional to 
$e^{2ky}$, is missing in Eq.(\ref{naive_radion}). 
Thus, the above naive ansatz must be corrected. 
The proper treatment of the radion mode is discussed 
in Refs.~\cite{CGR,CGK} in non SUSY case, and  
Ref.~\cite{BNZ} discusses this issue in the context of 5D pure SUGRA. 

In this talk, we will explain our work~\cite{AS2}, which has derived 
the 5D superspace action including 
the radion superfield and clarified its couplings to the matter superfields. 
Our work corresponds to an extension of Ref.~\cite{BNZ} 
including the matter superfields. 
The strategy of our work is as follows. 
First, we will identify how the radion mode appears in the superspace action. 
Then, we will promote it to a chiral superfield.

\section{Radion mode in the superspace action}
In our previous work~\cite{AS}, we have derived the 5D superspace action 
directly from 5D SUGRA action by fixing all the gravitational fields to their 
vacuum expectation values (VEVs).\footnote{
The full 5D superconformal gravity action is expressed in the language of 
the 4D superconformal gravity in Ref.~\cite{CST}. 
} 
\bea
 \vev{\ge{m}{\udl{n}}} \eql e^{\sgm(y)}, \nonumber\\
 \vev{\ge{y}{4}} \eql 1, \nonumber\\
 \vev{\psi_\mu} \eql 0, 
\eea
where $\ge{\mu}{\udl{\nu}}$ and $\psi_\mu$ are the f\"{u}nfbein and 
the gravitino, respectively. 
The function~$\sgm(y)$ is the warp factor, and $\sgm(y)=-ky$ 
in the case that the backreaction of the radius stabilizer field on the metric 
is neglected.\footnote{We will neglect such backreaction on the metric 
in the following.}
Since the radion mode originally belongs to the 5D gravitational multiplet, 
we have to modify the above treatment of the gravitational fields. 
To introduce the radion mode, 
we will replace the VEVs of f\"{u}nfbein with the radion-dependent 
functions~$F$ and $G$ as follows. 
\bea
 \sgm(y) &\to& F(b(x),y), \nonumber\\
 \vev{\ge{y}{4}} &\to& G(b(x),y), \label{replace}
\eea
where $b(x)$ is the radion field. 

There are two conditions that the functions~$F$ and $G$ must satisfy. 
The first condition comes from the requirement that 
our superspace action reproduces the correct 5D SUGRA action. 
To see this, let us take a kinetic term of a chiral superfield~$\Phi$ as an example. 
If we fix the gravitational fields to their VEVs, such kinetic term can be 
written as 
\bea
 \cL_{\rm kin} \eql e^{2\sgm}\vev{\ge{y}{4}}\int\dr^4\tht\;\bar{\Phi}\Phi 
 \nonumber\\
 \eql e^{2\sgm}\vev{\ge{y}{4}}\eta^{mn}\left\{-\frac{\bar{\vph}\der_m\der_n\vph}{4}
 -\frac{\der_m\der_n\bar{\vph}\vph}{4}\right. \nonumber\\
 &&\hspace{20mm}\left.+\frac{\der_m\bar{\vph}\der_n\vph}{2}+\cdots\right\} 
 \nonumber\\
 \!\!\!&\Rightarrow\!\!\!& 
 e^{2\sgm}\vev{\ge{y}{4}}\eta^{mn}\der_m\bar{\vph}\der_n\vph+\cdots, 
\eea 
where $\vph$ is the scalar component of $\Phi$. 
Note that we have performed the partial integral at the last step. 
After the replacement~(\ref{replace}), the prefactor~$e^{2\sgm}\vev{\ge{y}{4}}$ 
becomes $e^{2F}G$, which generically has a nontrivial $x$-dependence 
through the radion field~$b(x)$. 
However, if $e^{2F}G$ depends on $x^m$, unwanted extra terms appears 
after the partial integral. 
To avoid the appearance of such terms, we have to impose the condition 
that $e^{2F}G$ is independent of $x^m$. 
Considering its background value, this condition can be written as 
\be
 2F+\ln G = 2\sgm.  \label{cstrt1}
\ee
In fact, if we impose this condition, the correct radion kinetic term 
is also reproduced. 
Thus, this is the necessary and sufficient condition for the superspace action 
to reproduce the correct SUGRA action. 

The second condition comes from the fact that 
the radion field should appear in the metric 
as if it is a modulus field. 
In other words, the bulk geometry should remain $\mbox{AdS}_5$ with 
a definite curvature~$k$ when we shift VEV of the radion field by constant. 
This condition can be written as 
\be
 G = -\frac{1}{k}\der_y F.  \label{cstrt2}
\ee

By solving the above constraints~(\ref{cstrt1}) and (\ref{cstrt2}), 
explicit function forms of $F$ and $G$ are determined as follows. 
\bea
 F \eql \frac{1}{2}\ln\brkt{e^{2\sgm}+\cI(b)},  \nonumber\\
 G \eql \frac{1}{1+e^{-2\sgm}\cI(b)}, \label{G_form}
\eea
where $\cI(b)$ is some function of only $b(x)$ and satisfies 
\be
 \cI(\vev{b})=0. 
\ee
In the following, we will choose it as $\cI(b)=\tl{b}\equiv b-\vev{b}$. 
In this case, the metric agrees with that of Ref.~\cite{BNZ}. 

Since the most familiar definition of the radion field is a proper length~$r(x)$, 
we will rewrite $b(x)$ in terms of $r(x)$. 
From its definition, the proper length is written as 
\bea
 r(x) \defa \frac{1}{\pi}\int_0^{\pi R}\dr y\; G(b(x),y)
 = \frac{1}{\pi}\int_0^{\pi R}\frac{dy}{1+e^{-2\sgm}\tl{b}}  \nonumber\\
 \eql R-\frac{1}{2k\pi}\ln\brkt{\frac{1+e^{2k\pi R}\tl{b}}{1+\tl{b}}}, 
\eea
or equivalently, 
\be
 \tl{b} = e^{-k\pi R}\frac{\sinh k\pi (R-r(x))}{\sinh k\pi r(x)}. 
\ee
By substituting this into Eq.(\ref{G_form}), the function~$G$ can be expressed 
in terms of $r(x)$. 

Using this $G$, we can express the bulk Lagrangian as follows. 
\bea
 \cL \eql \cL_{\rm kin}^{\rm rad}
 +\brc{\int\dr^2\tht\;\frac{1}{4}G_{\rm c}\cW\cW+\hc} \nonumber\\
 &&+e^{2\sgm}\int\dr^4\tht\;G^{-2}\brkt{\der_y V+i\Phi_S-i\bar{\Phi}_S}^2 
 \nonumber\\
 &&-e^{2\sgm}\int\dr^4\tht\;\left\{2M_5^3\brkt{\bar{\Sgm}\Sgm}^{\frac{3}{2}}
 \right.\nonumber\\
 &&\hspace{20mm} \left.
 -G^{\frac{3}{2}}\brkt{\bar{H}e^{2gV}H+\bar{H}^C e^{-2gV}H^C}\right\} \nonumber\\
 &&+e^{3\sgm}\left\{\int\dr^2\tht\; H^C\brkt{\frac{\lrder_y}{2}+mG_{\rm c}
 -2ig\Phi_S}H\right. \nonumber\\
 &&\hspace{40mm} \left. +\hc\right\}, \label{r_dep_action}
\eea
where $M_5$ is the 5D Planck mass, $g$ is a gauge coupling, and  
$\cW$ is the superfield strength of the vector superfield~$V$.  
The chiral superfields~$(H,H^C)$ form a hypermultiplet, 
and $\Sgm$ and $\Phi_S$ are the 5D compensator superfield 
and the gauge scalar superfield, respectively.  
The complex quantity~$G_{\rm c}$ and the radion kinetic 
term~$\cL_{\rm kin}^{\rm rad}$ are defined as
\bea
 G_{\rm c} \defa G-i\kp W_y^0, \label{def_Gc} \\
 \cL^{\rm rad}_{\rm kin} \defa \frac{3M_5^3(k\pi)^2}{16}(1-e^{-2k\pi R})^2
 \frac{e^{-2\sgm}G^2(r)}{\sinh^4 k\pi r} \nonumber\\
 &&\times\eta^{mn}\der_m r\der_n r, \nonumber\\
\eea
where $\kp\equiv 1/M_5$ and $W^0_\mu$ is the graviphoton field.

\section{Promotion to the radion superfield}
In order to obtain the desired superspace action, 
we will promote the radion field~$r(x)$ in Eq.(\ref{r_dep_action}) 
to a superfield. 
Here, we will define a complex scalar~$\tau$ as 
\be
 \tau \equiv r+i\kp w, 
\ee
where $w(x)$ is a gauge-invariant Wilson line of the graviphoton,\footnote{
Here we will assume that $\vev{W_y^0}=0$. } 
\be
 w \equiv \frac{1}{\pi}\int_0^{\pi R}\dr y \; W_y^0. 
\ee
Then, we can easily check that the kinetic term for $\tau$ becomes 
the K\"{a}hler form. 
This fact suggests that $r(x)$ should be associated with $w(x)$ 
in the form of $\tau(x)$. 
For example, a complex quantity~$G_{\rm c}$ appearing in Eq.(\ref{r_dep_action}) 
should be interpreted as 
\be
 G_{\rm c} = G(\tau) 
 = \brc{1+e^{-2\sgm(y)}e^{-k\pi R}\frac{\sinh k\pi(R-\tau)}{\sinh k\pi\tau}}^{-1}. 
\ee
Then, from Eq.(\ref{def_Gc}), $G$ and $W_y^0$ are identified as 
\be
 G \equiv \Re\:G(\tau), \;\;\; W_y^0 \equiv -M_5\Im\:G(\tau). 
\ee

Now we will promote the complex scalar~$\tau$ to a chiral superfield~$T$. 
Namely, $G_c$ and $G$ in Eq.(\ref{r_dep_action}) are promoted as 
\bea
 G_{\rm c} &\to& G(T) =\brc{1+e^{2ky}e^{-k\pi R}\frac{\sinh k\pi(R-\tau)}
 {\sinh k\pi T}}^{-1}, \nonumber\\
 G &\to& G_{\rm R}\equiv \Re\:G(T).  \label{def_GT}
\eea

As a result, the desired superspace Lagrangian becomes 
\bea
 \cL \eql \brc{\int\dr^2\tht\;\frac{1}{4}G(T)\cW\cW+\hc} \nonumber\\
 &&+e^{2\sgm}\int\dr^4\tht\; G_{\rm R}^{-2}\brkt{\der_y V+i\Phi_S-i\bar{\Phi}_S}^2
 \nonumber\\
 &&+e^{2\sgm}\int\dr^4\tht\; G_{\rm R}^{\frac{3}{2}}
 \brkt{\bar{H}e^{2gV}H+\bar{H}^C e^{-2gV}H^C} \nonumber\\
 &&+e^{3\sgm}\left\{\int\dr^2\tht\;H^C\brkt{\frac{\lrder_y}{2}+mG(T)-2ig\Phi_S}H
 \right.\nonumber\\
 &&\hspace{60mm}\left.+\hc\right\} \nonumber\\
 &&-e^{2\sgm}\int\dr^4\tht\; 3M_5^3\ln G_{\rm R} \nonumber\\
 &&+\sum_{\vtht^*=0,\pi}\cL^{(\vtht^*)}_{\rm brane}\dlt(y-R\vtht^*), 
\eea
where $\cL^{(\vtht^*)}_{\rm brane}$ is a brane localized Lagrangian 
at $y=R\vtht^*$ given by 
\bea
 \cL^{(\vtht^*)}_{\rm brane} \eql \brc{\int\dr^2\tht\;f^{(\vtht^*)}_{AB}
 \cW^A\cW^B+\hc} \nonumber\\
 &&-e^{2\sgm}\int\dr^4\tht\;G_{\rm R}^{-1}\exp\brc{-K^{(\vtht^*)}(S,\bar{S},U)}
 \nonumber\\
 &&+e^{3\sgm}\brc{\int\dr^2\tht\;G_{\rm R}^{-\frac{3}{2}}(T)P^{(\vtht^*)}(S)+\hc}.
 \nonumber\\
\eea
Here we have assumed that the background preserves $\cN=1$ SUSY and 
dropped the compensator superfield. 
$\cW^A$ is a superfield strength of $U^A$. 
The chiral superfield~$S$ and the vector superfield~$U^A$ 
in $\cL^{(\vtht^*)}_{\rm brane}$ can be either brane-localized superfields or 
brane-induced superfields from the bulk superfields.

\section{Supersymmetric radius stabilization}
Finally, we will demonstrate the radius stabilization in a model proposed 
in Ref.~\cite{MO}, which corresponds to a supersymmetric extension of 
the Goldberger-Wise mechanism \cite{GW}. 

The stabilization sector consists of a hypermultiplet~$(H,H^C)$ 
with a bulk mass $m$. 
The following tadpole superpotentials are introduced at both boundaries. 
\be
 P^{(0)}=J_0H, \;\;\; P^{(\pi)}=-J_\pi H, 
\ee
where $J_0$ and $J_\pi$ are real constants. 

Then, we can calculate the radion potential from our superspace action 
obtained in the previous section. 
\bea
 V_{\rm rad}(\tau) \eql 
 \frac{\abs{G^{-\frac{\gm}{2k}}(\tau,y=0)}^2\cdot
 \abs{J_0-J_\pi e^{-\gm\pi\tau}}^2}
 {4\int_0^{\pi R}\dr y\;e^{(k-2m)y}G_{\rm R}^{\frac{3}{2}}
 \abs{G^{-\frac{m}{2k}}(\tau)}^2} \nonumber\\ 
 &&+\cO(l^4), 
\eea
where $\gm\equiv \frac{3}{2}k+m$ and $l\equiv \kp^{3/2}\abs{J_\pi}$. 
Here we have restricted the section of $h=0$, where $h$ is 
the scalar component of $H$. 
From this potential, we can easily see that the radius~$\vev{r}(=\Re\vev{\tau})$ 
is certainly stabilized to a finite value 
\be
 \vev{r}=\frac{\ln(J_\pi/J_0)}{\gm\pi}. 
\ee

By differentiating this potential with respect to $\tau$, we can also calculate 
the radion mass as 
\bea
 m_{\rm rad}^2 \!\!\eql \!\!\frac{l^2k^2}{6}\brkt{1-\frac{2m}{k}}
 \brkt{\frac{3}{2}+\frac{m}{k}}^2 \nonumber\\
 &&\times\frac{e^{-2k\pi R}(1-e^{-2k\pi R})}{1-e^{-(k-2m)\pi R}}+\cO(l^4). 
\eea
This completely agrees with the result obtained by directly solving 
the equation of motion. 
This agreement supports the validity of the $T$-dependence of 
our superspace action.

\section{Summary}
We have derived 5D superspace action including the {\it dynamical} radion 
superfield, and clarified its couplings to the bulk and the boundary matter 
superfields. 
Our result is obtained in a systematic way based on the superconformal 
formulation of 5D SUGRA in Ref.~\cite{KO}. 

The $T$-dependence of our action is different from that of Ref.~\cite{MP}, 
which is based on the naive ansatz~(\ref{naive_radion}). 
Especially, the marked difference appears 
in the couplings between $T$ and the K.K. modes of the matter superfields. 

Note that we cannot redefine $G(T)$ as a single chiral superfield 
by holomorphic redefinition of the superfield 
because $G(T)$ has an explicit $y$-dependence. 
(See Eq.(\ref{def_GT}).)
Thus, the 4D effective K\"{a}hler potential has a quite complicated form 
in the case of the warped geometry. 
In contrast, the $T$-dependence becomes 
greatly simplified in the flat spacetime (\ie, $k=0$). 
The $y$-dependence of $G(T)$ disappears and 
\be
 G(T)=\frac{T}{R}. 
\ee
The 4D effective K\"{a}hler potential in this case becomes the following 
no-scale form up to a constant. 
\be
 K^{(4)}_{\rm rad}(T,\bar{T}) = -3M_P^2\ln\brkt{T+\bar{T}}, 
\ee
where $M_P=(\pi RM_5^3)^{1/2}$ is the 4D Planck mass. 

Here, we have assumed that the background preserves $\cN=1$ SUSY, 
and dropped the dependence of the compensator superfield. 
However, it plays an important role when we consider the SUSY breaking effects. 
Thus, our next task is to extend our result including the compensator superfield. 
Note that what appears in the 4D effective action is 
the {\it 4D} compensator superfield. 
Although the superconformal formulation of 5D SUGRA has 
a compensator multiplet, they are {\it 5D} fields. 
Since the compensator multiplet is not dynamical, 
we cannot mode-expand its component fields into the K.K. modes 
in the ordinary manner. 
In fact, the $F$-terms of the compensator and the radion superfields 
are closely related to each other. 
Therefore, the result we have derived here
provides an important hint for identifying 
the dependence of the 5D superspace action on the 4D compensator superfield. 
The research along this line is now in progress.


\vspace{3mm}

\begin{center}
{\bf Acknowledgments}
\end{center}
 This work was supported by KRF PBRG 2002-070-C00022 (H.A.), 
 and the Astrophysical Research Center for the Structure 
 and Evolution of the Cosmos (ARCSEC) funded by the Korea Science and Engineering 
 Foundation and the Korean Ministry of Science (Y.S.). 
 Y.S. is supported by the Japan Society for the Promotion of Science for 
 Young Scientists (No.0509241). 





\begin{thebibliography}{9}
 \bibitem{RS} L.~Randall and R.~Sundrum, \PRL{83} (1999) 3370 
  [{\tt hep-ph/9905221}]. 

 \bibitem{GW2} W.D.~Goldberger and M.B.~Wise, \PL{B475} (2000) 275 
  [{\tt hep-ph/9911457}]. 

 \bibitem{CGRT} C.~Cs\'{a}ki, M.~Graesser, L.~Randall and J.~Terning, 
  \PR{D62} (2000) 045015 [{\tt hep-ph/9911406}]. 

 \bibitem{CGR} C.~Charmousis, R.~Gregory, and V.A.~Rubakov, 
  \PR{D62} (2000) 067505 [{\tt hep-th/9912160}]. 

 \bibitem{CGK} C.~Cs\'{a}ki, M.L.~Graesser, and G.D.~Kribs, 
  \PR{D63} (2001) 065002 [{\tt hep-th/0008151}]. 

 \bibitem{BNZ} J.~Bagger, D.~Nemeschansky and R.~Zhang, 
  \JH{0108} (2001) 057 [{\tt hep-th/0012163}]. 

 \bibitem{AS2} H.~Abe and Y.~Sakamura, \PR{D71} (2005) 105010 
  [{\tt hep-th/0501183}]. 

 \bibitem{AS} H.~Abe and Y.~Sakamura, \JH{0410} (2004) 013 
  [{\tt hep-th/0408224}]. 

 \bibitem{CST} F.~Paccetti Correia, M.G.~Schmidt and Z.~Tavartkiladze, 
  \NP{B709} (2005) 141 [{\tt hep-th/0408138}]. 

 \bibitem{MO} N.~Maru and N.~Okada, \PR{D70} (2004) 025002 
  [{\tt hep-th/0312148}]. 

 \bibitem{GW} W.D.~Goldberger and M.B.~Wise, \PRL{83} (1999) 4922 
  [{\tt hep-ph/9907447}]. 

 \bibitem{KO} T.~Kugo and K.~Ohashi, \PTP{105} (2001) 323 
  [{\tt hep-ph/0010288}]; \PTP{108} (2002) 203 
  [{\tt hep-th/0203276}]. 

 \bibitem{MP} D.~Marti and A.~Pomarol, \PR{D64} (2001) 105025 
  [{\tt hep-th/0106256}].
\end{thebibliography}

\IfFileExists{\jobname.bbl}{}
 {\typeout{}
  \typeout{******************************************}
  \typeout{** Please run "bibtex \jobname" to optain}
  \typeout{** the bibliography and then re-run LaTeX}
  \typeout{** twice to fix the references!}
  \typeout{******************************************}
  \typeout{}
 }

\end{document}